# Towards Commodity, Web-Based Augmented Reality Applications for Research and Education in Chemistry and Structural Biology

Luciano A. Abriata*

École Polytechnique Fédérale de Lausanne, Lausanne, Switzerland

luciano.abriata@epfl.ch - http://labriataphd.altervista.org/

## ABSTRACT

This article reports prototype web apps that use commodity, open-source technologies for augmented and virtual reality to provide immersive, interactive human-computer interfaces for chemistry, structural biology and related disciplines. The examples, which run in any standard web browser and are accessible at https://lucianoabriata.altervista.org/jsinscience/arjs/armodeling/ together with demo videos, showcase applications that could go well beyond pedagogy, *i.e.* advancing actual utility in research settings: molecular visualization at atomistic and coarse-grained levels in interactive immersive 3D, coarse-grained modeling of molecular physics and chemistry, and on-the-fly calculation of experimental observables and overlay onto experimental data. From this playground, I depict perspectives on how these emerging technologies might couple in the future to neural network-based quantum mechanical calculations, advanced forms of human-computer interaction such as speech-based communication, and sockets for concurrent collaboration through the internet -all technologies that are today maturing in web browsers- to deliver the next generation of tools for truly interactive, immersive molecular modeling that can streamline human thought and intent with the numerical processing power of computers.

## 1 INTRODUCTION

Virtual reality (VR) and augmented reality (AR)[1] have enormous potential in assisting scientific research and education, especially in disciplines that deal with abstract objects, objects much smaller or larger than human dimensions, or objects that are hard to acquire and handle due to high costs, limited availability, fragility, etc.[2], [3] Chemistry is one such discipline where AR and VR have been attributed high educational and exploratory potentials [4]–[9], although actual reach, user acceptance and educational effects still need deeper evaluation.[10]–[14] Arguably, some of these limitations arise from the complex software setups involved in most current applications, which combine different modules to take camera feeds, analyze markers, calculate molecular data and render overlaid graphics; and also from the need of specialized hardware in most available solutions. As I intend to demonstrate here, these and other problems are alleviated when client side web technologies for webcam-based marker tracking are utilized.

More specifically, in this article I demonstrate, through examples from the domains of chemistry, biophysics and structural biology, that client side web-based technologies have matured enough to support complete scientific applications based on AR and VR through commodity software and hardware in commonplace web browsers, which educators and researchers can already put into use today. Being based entirely on open, community-contributed client-side software for building interactive web content, the web apps presented here are on one hand easy to deploy, profiting from several ready-to-use libraries and APIs, and on the other hand freely and fully accessible to users through standard web browsers in multiple devices such as smartphones, tablets and computers, without requiring any installation. This and other advantages of web app technologies are discussed in [15], [16].

Moreover, the presented prototype web apps introduce novel features that bring interactivity and thus utility beyond visualization. Previous works have focused mainly on visualization largely overlooking molecular dynamics, reactivity and information overlay, except for some notable works like [17], [18]; but as the examples show, current technologies allow for efficient in-browser calculations of molecular mechanics, calculations of experimental observables and real-time comparison to actual experimental data, plus enhanced human-computer interaction including speech-based commands, and even more exciting possibilities as discussed.

## 2 METHODS
### 2.1 Software
From the software side, the core of the prototype applications presented here relies in community-contributed in-browser (*i.e.* JavaScript-based) solutions for AR, VR, simulation of particle physics, speech recognition and data plotting (Table 1), put together and extended with code written *ad hoc*. The reason for using community-contributed tools is to keep code open, simple and accessible to interested students and teachers. Figure S1 at the end of the article shows sample code for one of the simplest examples.

Specifically, AR and VR are achieved through A-Frame, an entity component system framework for Three.js that simplifies creation of 3D and virtual scenes in the web. Marker tracking is achieved through AR.js developed by J. Etienne, handling rendering of Three.js objects through A-Frame. Solid body physics are handled through Cannon.js developed by S. Hedman and wrapped into HTML by D. McCurdy. Calculations specific to the problems of interest are coded *ad hoc* in JavaScript. The annyang speech recognition library by T. Ater is used to process spoken commands, and Google Charts is used for displaying data plots. All relevant URLs pointing to these tools are given in Table 1.

In all examples, atoms and beads (beads being groups of atoms in coarse-grained descriptions of molecules) are treated as spheres, displayed as A-Frame <a-sphere> components coded directly in the HTML. Residue beads for single-bead-per-residue coarse-grained description of proteins are centered at CA atoms and their radii are calculated as the cubic root of the corresponding amino acid volumes as compiled in [19]. Bonds between atoms or residues are displayed as <a-tube> components registered by Don McCurdy's A-Frame-extras. Sphere and tube tags are affected by an id that gives handlers for subsequent operations, and are placed inside <a-marker> tags, which are components registered by the A-Frame-compatible AR.js to handle AR markers.

The examples provided here use the Hiro and Kanji markers built-in inside AR.js (Figure S2). One or more marker tags are located inside one <a-scene> tag, which includes also a <a-entity camera> tag to locate a camera for rendering. Style definitions at the beginning of the HTML file setup the video feedback and display. Video might need to be horizontally or vertically flipped depending on the relative orientation of the user, the webcam, and the screen or projection in the different hardware setups (below and Figure S3).

### 2.2 Hardware
The web apps presented here were tested on laptops with built-in webcams running Windows (using Firefox and Chrome web browsers), Linux (only Firefox tested) and Mac (Safari web browser) operating systems, (ii) smartphones and tablet devices with built-in cameras running Android (Chrome web browser), and desktop computers with plugged webcams and running Linux Fedora and Windows operating systems 6 (Edge, Firefox and Chrome browsers).

Different hardware platforms offer different possibilities for layout (Figure S3). For example, a smartphone can be used inside cardboard-based goggles to achieve a more immersive experience, and multi-person AR/VR. On the other hand, a webcam mounted on a desktop computer can offer better graphics performance and calculation power, and space for user's arms resting on the table, but is less comfortable for concurrent use by multiple persons.

**Table 1.** Client-side libraries, APIs and components used to build the web apps reported here.

| Tool | URL |
|---|---|
| AR.js | https://github.com/jeromeetienne/AR.js |
| A-Frame | https://aframe.io/ |
| Three.js | https://threejs.org/ |
| Cannon.js through A-Frame physics | http://www.cannonjs.org/ and https://github.com/donmccurdy/aframe-physics-system |
| Annyang | https://github.com/TalAter/annyang |
| Google Charts | https://developers.google.com/chart/ |

## 3 PROTOTYPE WEB APPS
This section presents sample web apps (compatible with all major web browsers in modern smartphones, tablets and computers) that introduce features of increasing complexity, organized in a first set that involves small molecules and a second set that involves proteins. Tables 2 and 3 list the URLs to the specific examples mentioned, also accessible together with other relevant resources and example videos at https://lucianoabriata.altervista.org/jsinscience/arjs/armodeling/. Users interested in developing their own HTML web apps can use the tools provided in that website under "Building tools", continuously updated as more complex examples are developed and new features added. When using these building tools, atomic coordinates must be provided in PDB format. PDB files for proteins can be downloaded from the Protein Data Bank and related resources [20] while small molecules can be built with Hack-a-Mol at https://chemapps.stolaf.edu/jmol/jsmol/hackamol.htm, a tool based itself on client-side web software like JSmol [21] and JSME [22].

## 3.1 Examples on small molecules

### 3.1.1 Simplest, HTML-only examples

The simplest web app example consists exclusively of HTML code (Table 2A). With it, the user drives a lysine side chain with the Hiro marker and a glutamate side chain with the Kanji marker. Each molecule is anchored to the center of its corresponding AR marker through its CA atom. Their protonation states correspond to neutral pH, so lysine is protonated hence its N atom (blue) is charged by +1, whereas glutamate is deprotonated hence its O atoms (red) bear a total charge of -1. These charges do not have any effect on this example, but they do in the examples B and C.

**Table 2.** Links to the example web apps about small molecules.

| Web app | URL |
|---|---|
| A) HTML-only small molecules | https://lucianoabriata.altervista.org/jsinscience/arjs/armodeling/smallmolsimplest.html |
| B) (A) plus clash detection | https://lucianoabriata.altervista.org/jsinscience/arjs/armodeling/smallmolclashdetection.html |
| C) (B) with phone vibration | https://lucianoabriata.altervista.org/jsinscience/arjs/arjs-phones.html |
| D) (B) with proton transfer | https://lucianoabriata.altervista.org/jsinscience/arjs/armodeling/smallmolprotontransfer.html |
| E) Diels Alder reaction | https://lucianoabriata.altervista.org/jsinscience/arjs/armodeling/smallmoldielsalder.html |
| F) Glucose enantiomers and isomers | Linear L- vs. linear D- glucose: https://lucianoabriata.altervista.org/jsinscience/arjs/armodeling/smallmolLglcDglc.html D-furanose vs. D-pyranose: https://lucianoabriata.altervista.org/jsinscience/arjs/armodeling/smallmolglcfurapyra.html |
| G) | https://lucianoabriata.altervista.org/jsinscience/arjs/armodeling/metmyoglobinfe3pcshift.html |

### 3.1.2 Adding chemistry and physics with JavaScript

Example B (Figure 1) features the same two molecules as example A, but includes simple JavaScript code to (i) adapt the positions of 10 yellow spheres that make up a dotted line connecting in space the lysine's N atom with one of the glutamate's O atoms; (ii) report the distance between these 2 atoms and the corresponding attractive electrostatic force in real time; and (iii) test and display clashes between any pair of atoms of the two molecules. The code for (i) is wrapped inside the dashedline.js file; whereas the code for (ii) and (iii) is located between <script> tags at the end of the HTML file. The distance, electrostatic and clash calculations are inside a setInterval() function that is executed every 200 ms and uses the id identifiers of the sphere tags to locate them.

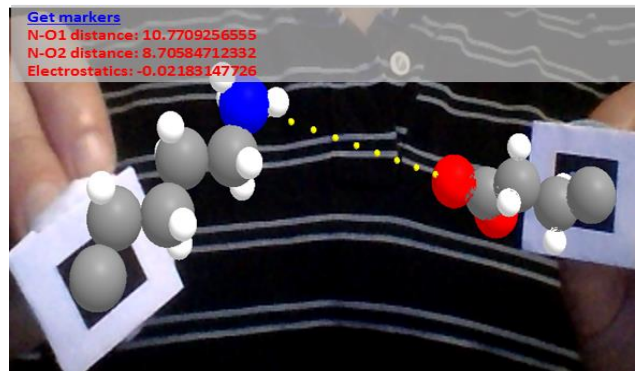

**Figure 1:** Example B from Table 2, two small molecules driven in space with AR markers, augmented with simple calculations in real time. The code for this example is shown in Figure S1.

The distance is calculated in Angstrom and includes a correction for a zoom factor that scales atom sizes and positions when the molecular coordinates are parsed into A-Frame HTML, to properly fit the screen (in this example a factor of 2). Clashes are detected as two spheres being within 3 Å, and displayed as semitransparent A-Frame spheres centered on the affected atoms.

Example C is the same as B but is more suited for phones, as it executes a 200 ms long vibration when a clash is detected. Also, its video display is flipped horizontally compared to the other examples, to allow use with commodity VR lenses such as Google Cardboard (example in Figure S3).

Another use of in-browser calculations is the simulation in real time of experimental observables at each configuration of the molecular system. Figure 2 (link G in Table 2) exemplifies this with the calculation of paramagnetically induced pseudocontact chemical shift and line broadening on a probe atom attached to one AR marker, as it is moved around the heme group of metmyoglobin with the other AR marker. This web app implements standard equations from the theory of paramagnetic nuclear magnetic resonance [23] fed on the fly with the corresponding polar coordinates, then simulates the spectrum including noise and displays it using Google Charts.

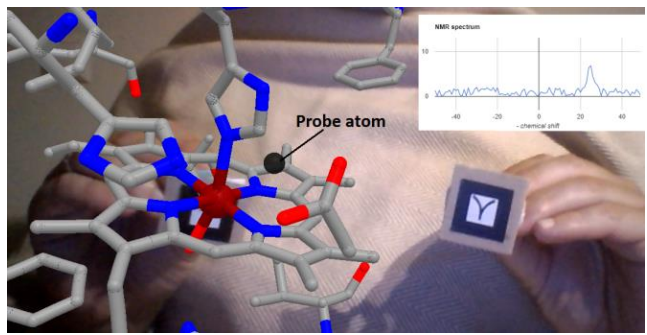

**Figure 2:** The active site of Fe(III) metmyoglobin on the Hiro marker, and a probe hydrogen atom on the Kanji marker. From the polar coordinates of the probe atom in the framework of the heme group, pseudocontact chemical shifts and line broadening are calculated and the corresponding paramagnetic spectrum is simulated, displayed with Google Charts.

### 3.1.3 Simulation of chemical reactivity and prospect for quantum-based calculations

Example D extends example B by incorporating a very simplistic version of distance-based proton transfer. As explained in 3.1.1, the lysine side chain is protonated and the glutamate side chain is deprotonated. Example D uses JavaScript to stochastically transfer one proton from lysine's N atom to one of the O atoms of glutamate if the lysine is protonated, or the other way around. Proton transfers occur only when the proton is within 2 Å of the receiving N or O atom, and jumps back and forth to reflect 70% time-averaged population of protonated lysine and 30% of protonated glutamate, to convey the feeling of different acidic constants. At distances longer than 2 Å but shorter than 3 Å the web app displays a yellow dotted line that represents a hydrogen bond between the potential receiver heavy atom and the involved proton.

The next example, E in Table 2, could allow teachers to illustrate stereoselectivity in the Diels-Alder reaction in interactive 3D. This reaction occurs between a dienophile and a conjugated diene. It takes place in a concerted fashion, so the side of the diene where the initial approach occurs defines the stereochemistry of the product. The web app in this example allows users to visualize all this in 3D as they approach a molecule of 1,3-cyclohexadiene held in the left hand (Hiro marker) and a molecule of vinyl chloride in the right hand (Kanji marker). As the two pairs of reacting C atoms approach simultaneously, the two new bonds gain intensity until the product is formed. The product formed in this reaction is by itself an interesting molecule to visualize and move around in 3D through AR, because it contains two fused six-membered rings which are often hard to grasp in 2D (Figure 3).

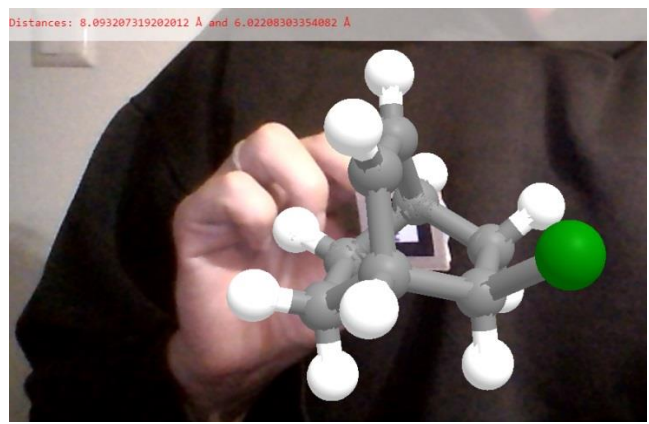

**Figure 3:** The bicyclic product of the simulated reaction in example E from Table 2, where carbon atoms are grey, hydrogens white and chlorine green.

There is a very important point regarding this section on simulated reactivity. Although these examples use merely artificial ways to visualize the proton transfer or Diels-Alder reactions, they are not based on actual quantum calculations. Such calculations are extremely slow to be incorporated into immersive experiences where energies need to be computed on the fly. However, novel machine learning methods that approximate quantum calculations through $>10^5$ times faster computations (like in [24]–[26]) could in a near future be coupled to AR/VR systems to interactively explore reactivity with realistic energy profiles in real time. Such tools would be useful not only for education but also for actual research, for example to interactively test the effect of chemical substituents on a reaction, estimate effects on spectroscopic observables, probe effects of structural changes on molecular orbitals, etc.

### 3.1.4 Other examples for small molecules

Visualization in 3D helps immensely to grasp the geometric details of complex small molecules and concepts about chirality, planarity and nonplanarity, rings, etc. The main website contains, under the "Small molecules" menu, other examples on top of those described here. For instance, example F from Table 2 allows exploration of 3D structures in molecules like glucose in its multiple forms, likely useful for pedagogical purposes.

**Table 3.** Links to example web apps dealing with protein systems.

| Web app | URL |
|---------|-----|
| A) Protein secondary structures | https://lucianoabriata.altervista.org/jsinscience/arjs/armodeling/secondarystructures.html |
| B) Standing alpha helix | https://lucianoabriata.altervista.org/jsinscience/arjs/armodeling/bigalphahelix.html |
| C) Two proteins in atomic detail | https://lucianoabriata.altervista.org/jsinscience/arjs/armodeling/ubiquitinandNESatomistic.html |
| D) Ubiquitin-UIM, 1 bead/res. | https://lucianoabriata.altervista.org/jsinscience/arjs/armodeling/ubiquitinuimsimplest.html |
| E) (D) with active restraints and contact force field | https://lucianoabriata.altervista.org/jsinscience/arjs/armodeling/ubiquitinuimforcefieldandvoice.html |
| F) (E) with real-time comparison of SAXS profile | https://lucianoabriata.altervista.org/jsinscience/arjs/armodeling/ubiquitinuimffvoicesaxs.html |
| G) Manual docking guided by contact data | https://lucianoabriata.altervista.org/jsinscience/arjs/armodeling/coevol_1qop.html |
| H) (D) with proteins connected through flexible linker | https://lucianoabriata.altervista.org/jsinscience/arjs/armodeling/ubiquitin-uim-cannon.html |

### 3.2 Examples on biological macromolecules

Biological macromolecules include large covalent polymers like proteins, nucleic acids and storage sugars, as well as non-covalent assemblies such as lipids forming membranes and complexes of different kinds of macromolecules, like ribonucleoproteins or integral membrane proteins.

In standard computational molecular modeling, macromolecules are represented either explicitly accounting for all their atoms (atomistic models), or by grouping atoms into beads, so-called coarse-grained models. Such grouping into beads decreases the number of particles to be simulated dramatically, and reduces the degrees of freedom of the system, thus overall reducing the mathematical complexity of the system [27].

Common coarse-graining schemes include grouping four heavy atoms and their attached hydrogens into a single bead, or even representing each unit of the polymer as a single sphere. So far the tools available at the website allow treating proteins either at atomistic resolution, by using the same building tool developed for small molecules, or at two coarse-grained levels, either 1 bead per residue (centered at CA atoms), or 2-to-4 beads per residue separating backbone from side chain to deliver more realistic shape than 1 bead per residue (similar to the MARTINI[28] coarse-graining scheme).

### 3.2.1 HTML-only AR for proteins at atomistic and coarse-grained descriptions

For small proteins and for sections of large proteins, atomistic representations are useful and can be well displayed even in smartphones (as is in fact the case of metmyoglobin advanced in Figure 2 as a small molecule case). Examples A and B from Table 3 use HTML only and were built from PDB structures using the same small molecule builder web app used in section 3.1 for small molecules, with the only remark that molecules were first centered at 0,0,0 in an external program, here VMD[29].

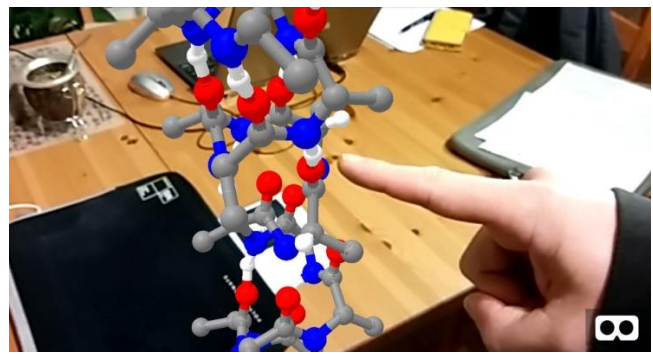

**Figure 4:** Example B from Table 3, a standing alpha helix as seen through an Android smartphone in mobile Chrome. H atoms are in white, O in red, N in blue and C in grey. The icon in the bottom right splits the screen in two for stereo projection in commodity AR lenses.

Example A from Table 3 displays on the Hiro marker a canonical alpha helix made up of consecutive alanine residues interrupted with a proline, and on the Kanji marker a beta-hairpin from an actual NMR structure (PDB ID 2k6v[30]), both in atomic detail including the hydrogen atoms (excluding side chains to better show the backbone hydrogen bonds). These examples allow 3D exploration of the two main kinds of secondary structures found in proteins, readily highlighting the strong i-i+4 hydrogen bonds in alpha helices, the trans-beta-sheet hydrogen bonds of antiparallel beta strands, the role of prolines as alpha helix breakers, and the twisted nature of beta hairpins. Example B (Figure 4) presents a standing helix that can be visualized from a Hiro marker lying on a flat surface, such that teachers and surrounding students could concurrently inspect from different angles. In such situations, a proper AR setup should allow multiple users to point in real space at objects in a way that all other users can see simultaneously, as exemplified in Figure S4.

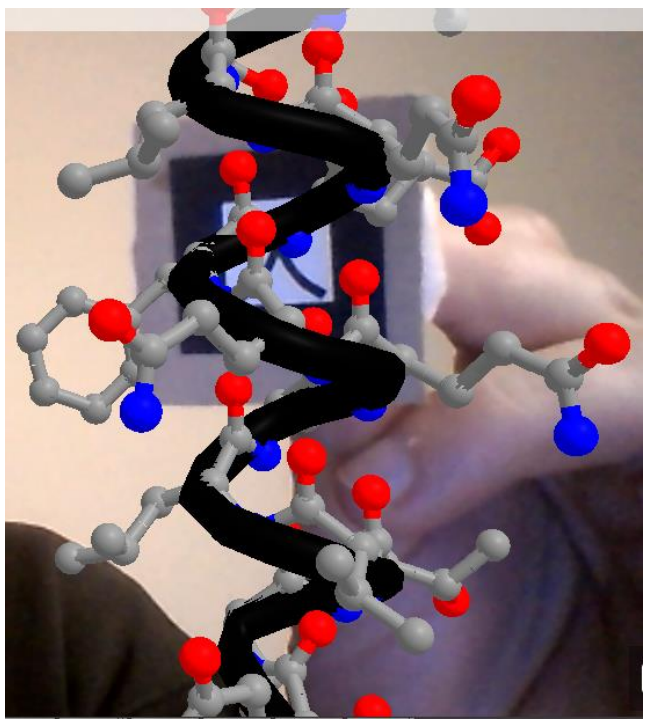

**Figure 5:** The nuclear export signal from example B in Table 3, modeled as an alpha helix, shows three hydrophobic residues on the left face and polar residues on the right.

Example C from Table 3 displays on the Hiro marker an NMR structure of ubiquitin (PDB ID 1d3z[31], model 1), and on the Kanji marker a helical model of the nuclear export signal of *Arabidopsis thaliana* Argonaute 1 protein [32] (Figure 5). Both visualizations include also a CA trace colored black, computed with another web app of the "Building tools" menu. For simplicity, the hydrogen atoms are omitted in these examples.

The ubiquitin example allows exploration of a small but complete protein structure, that could be further augmented to show residue names and details like non-covalent interactions, leading to a complete interactive 3D resource on protein structure. On the other hand, the example about the helical nuclear export signal readily shows the functionally relevant amphipathic nature of these protein motifs.

Example D from Table 3 deals again with ubiquitin, this time represented at a coarse-grained level of 1 bead per residue, and a ubiquitin-interacting motif (UIM) of helical structure, also represented at 1 bead per residue. Ubiquitin is displayed and driven by the Hiro marker, and UIM by the Kanji marker. The beads are colored according to amino acid type (blue = positively charged, red = negatively charged, green = polar uncharged, and grey = hydrophobic). They are semitransparent except for one bead from each protein that is opaque; these two opaque beads define a restraint coordinate for non-covalent interaction depicted as a yellow dashed line that connects them through space as shown for a related example in Figure 6.

### 3.2.2 Restraints, force fields and match to experimental data for interactive, integrative modeling of proteins

Example D just introduced is the starting point for more complex cases where the coordinates of ubiquitin and UIM proteins are fed into calculations (coded in JavaScript) useful for molecular modeling, experiment analysis and prediction, etc.

The first of these examples, E in Table 3, incorporates a rudimentary residue-resolution force field for clash detection and a molecule-resolution force field to apply restraints along a binding coordinate. These calculations involve iteration through all possible pairs of beads representing each protein's residues, computing distances and updating display (molecule positions and bead opacity) accordingly. Clashes between beads (i.e. residues) of the two proteins are manifested as more opaque colors. Restraints for binding are applied on UIM, by gradually shifting it in the coordinate of the Kanji marker.

Example F (Figure 6) goes one step further into actual uses that these technologies could have in structural biology and molecular modeling.

Building from example E, it incorporates on-the-fly calculation of small angle X-ray scattering (SAXS) profiles for the protein mixture and compares this profile to an experimental reference, offering a way to interactively test possible docking poses that are compatible with the experimental data. The SAXS profile calculation is based on the Debye formula iterated through pairs of residues instead of iterating through all atoms as the full equation requires, for simplicity and speed. It should be noted though that realistic SAXS profiles of high utility can be achieved by using just two scattering bodies per residue[33], a finding that enables realistic, detailed, real-time SAXS calculations of true utility in the near future.

In examples E and F, since the user's hands are both busy handling the AR markers, control of the clash-detection force field, restraint application, and SAXS calculations is achieved through spoken commands in English language, processed through the browser's speech recognition API. The possibility of incorporating such feature illustrates the superb integration capacity of libraries for client-side scripting in web browsers.[16]

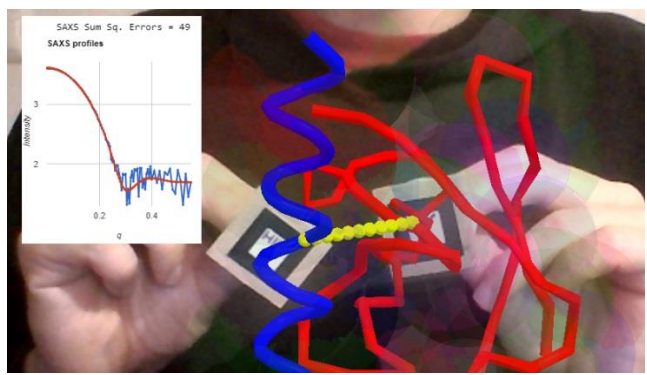

**Figure 6:** Ubiquitin (red trace) and ubiquitin-interacting motif (blue) driven in 3D with two AR markers, as the web app computes the predicted SAXS profile and displays it overlaid on top of an experimental (simulated with noise) profile, together with a metric of the fit quality.

Another example, G from Table 3, shows how AR can help to explore residue-residue contact data (Figure 7). This information essentially points at pairs of residues that are in contact, for example in the structure of a protein-protein complex. The example shows contacts predicted from coevolution analysis large sequence alignments, a bioinformatic tool now increasingly used to model protein structures and interactions.[34] The example in Figure 7 consists of coevolution data for proteins in chains A and B of PDB ID 1QOP, taken as is from [35] (and available at http://gremlin.bakerlab.org/cplx.php?uni_a=1QOP_A&uni_b=1QOP_B). In this example, each protein is driven by one marker, and the predicted contacts are overlaid as dotted lines connecting the intervening pairs of residues. These lines are colored green, olive and red according to the coevolution score, and their widths reflect in real time the distance between pairs of residues, supposed to be minimal when contact are satisfied.

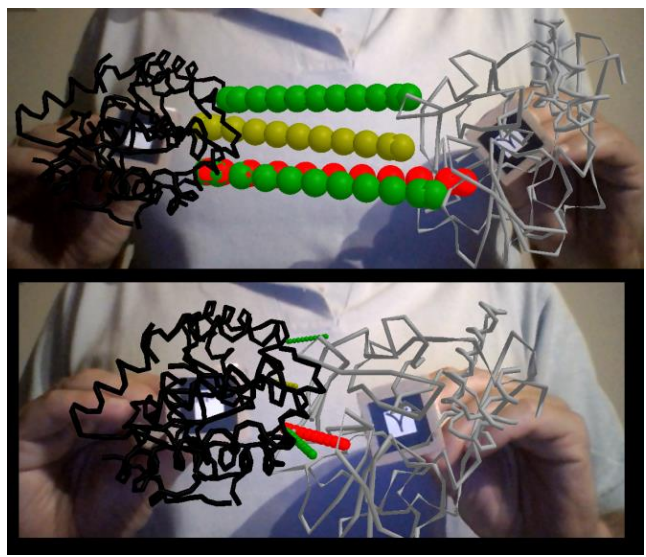

**Figure 7:** Interactive exploration of contacts predicted between two proteins (here from coevolution data) before (top) and after (bottom) manual docking. This example is 1QOP from [35], where contacts predicted with high score are colored green, contacts of intermediate confidence are olive, and the first contact of low probability is shown red (as taken from the original data). The thickness of the contact lines indicates distance, such that thin lines indicate the residues are close in space. Notice how the red contact, which has low probability, remains thicker than the well-scored contacts (green) upon manual docking.

### 3.2.3 Realistic handling of flexible interdomain linkers in proteins

Molecular motions, or "dynamics" or "flexibility", are essential for protein function and regulation, and hence must be accounted for in advanced modeling platforms. Broadly speaking, molecular motions range from restricted, coordinated structural changes in ordered domains, to extended dynamics in highly flexible regions. Correctly accounting for dynamics across that vast range of time and size scales is an active goal of research in molecular dynamics simulation methods, and requires complex force fields and molecular dynamics engines. Very encouragingly, a recent article showed that it is feasible to code such simulation methods in JavaScript [36] achieving near-native performance. For AR and VR applications, though, a number of yet unclear modifications are required in standard molecular dynamics engines to make them compatible with user-driven moves of the molecules; thus this is an area that requires research and probably a series of alternative solutions tailored to different problems.

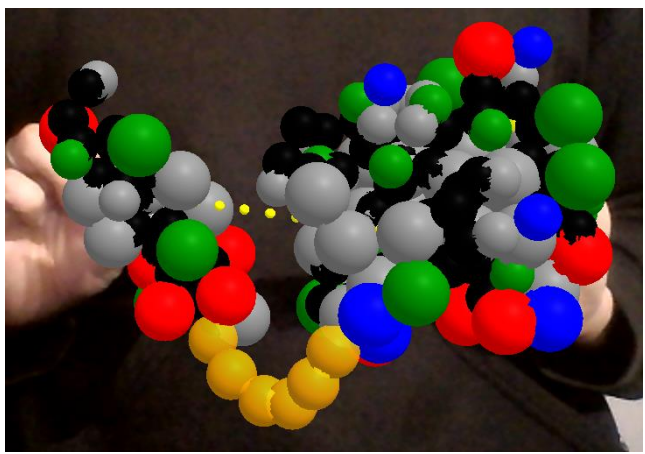

**Figure 8:** Example H from Table 3, ubiquitin and ubiquitin-interacting motif modelled as 2-4 beads per residue, colored by physiochemical properties (grey=hydrophobic, red=negative, blue=positive, green=polar uncharged) driven in 3D with two AR markers and with their termini connected through a flexible linker of six backbone-sized beads (orange).

The examples proposed here consider only the dynamics of very flexible linkers within the very coarse approximation that beads describing amino acids are rigid spheres connected by fixed distances. In the examples, such flexibility is handled by the Cannon.js software, wrapped through the A-Frame-physics extension. Example H is based in ubiquitin and its interacting partner as in D but in a fictitious scenario where they are covalently connected by 6 amino acids that form a flexible linker joining ubiquitin's C-terminus to the partner's N-terminus (Figure 8). Here, each globular domain (corresponding to ubiquitin or the interacting helix) is modelled with 2- to 4-beads per residue, while the flexible linker is modelled as backbone beads only. Such very simple model can help to answer complex questions related to the separation of the anchor points and the maximal extension of the linker when straight, for example: How far can the two proteins go with the given linker? Can the interacting partner be docked to satisfy the restraint yet allow for a relaxed configuration of the linker? Such investigations are in turn assisted by on-the-fly estimation of entropy associated to given extensions of the linkers, calculated here from a worm-like model [37]. More complex settings could in turn apply more elaborate physics descriptions, compute experimental observables to match in real time like as in the SAXS example F, etc.

## 4 DISCUSSION AND OUTLOOK

Achieving smooth and truly useful immersive chemistry (and biology) is one of the key "grand challenges for the simulation of matter in the 21st century" recently discussed by Aspuru-Guzik et al. [38] After around two decades of works slowly introducing AR and VR in chemistry, this year three very inspiring pieces of literature were published which show the real potential of modern, multiuser AR/VR in immersive molecular modeling and visualization. Goddard et al recently published a rich discussion about these technologies in the context of their own computer programs, including the widely used Chimera,[39] surveying advantages and disadvantages and presenting several interesting case studies.[13] O'Connor et al presented a VR-based system for multiuser handling of molecules in space, which even accounts for molecular mechanics through a standard force field.[40]

Although the hardware required by current solutions for AR/VR is much more accessible than a decade ago, it is still not what one could consider inexpensive and widely available across the world. Instead, the setup as introduced here requires only webcam-enabled devices, and as such works on any standard type of computer, laptop, tablet or smartphone. This simplicity also holds at the level of software, as the technology used here relies entirely on web browser-based code, i.e. HTML, CSS and JavaScript, using on top WebGL, free speech recognition engines, and other technologies built into web browsers. In fact, the reader could right now print a Hiro and Kanji AR marker from Figure S2, access the webpage at https://lucianoabriata.altervista.org/

 and test the examples him/herself.

The examples introduced here hence advance a potential for open, inexpensive, web-based AR and VR technologies in education and research in (bio)chemistry and related disciplines. In education, such tools could replace/complement tangible modeling kits further allowing virtually unlimited numbers and kinds of atoms, as well as augmenting models with additional information such as forces, charges, electron clouds and orbitals, data facts, etc. In research, such tools can help to visualize and probe molecular structure, simulate expected outcomes of experiments and test models and simulated data against experimental data, etc., everything through intuitive cues and fluent human-computer interactions.

Being these web apps so openly accessible, the next step is to develop content that teachers, students and researchers can put into use, to enable proper large-scale evaluations of the actual impact in learning and thinking processes. If successful, the near future will likely see these tools blended with modern molecular simulations and visualization methods, resulting in fully-fledged programs for deep immersive, interactive molecular modeling experiences that overcome the severe limitations of traditional software based on screens, mice and keyboards. Furthermore, happening all inside web browsers will seamlessly enable online concurrent collaboration among multiple peers as well as between students and teachers, as exemplified in [41]. Further adding haptic devices for force feedback, speech-based voice commands, numerical simulations and intelligent data and text mining –all technologies already built into web browsers- will bring research and education to the next level where human thought and intent couple with computer power to get the best out of both worlds.

## ACKNOWLEDGEMENTS

I acknowledge all the developers of the client-side web tools listed in Table 1, as well as the very helpful communities who contribute to their improvement, bug detection and correction, documentation, and online help.

I also acknowledge numerous researchers from the chemistry, physics and biology communities, who have contributed several useful ideas and suggestions, inspiring some of the presented examples.

**SUPPORTING INFORMATION AVAILABLE FROM NEXT PAGE**

# SUPPORTING INFORMATION

```html
<!DOCTYPE html>
<script src="../aframe.min.js"></script>
<script src="../aframe-ar.js"></script>
<script src="../client.js"></script>

<html>
<body style='margin : 0px; overflow: hidden; font-family: Monospace;'>
<style>
  canvas  { -moz-transform: scale(-1, 1);  -webkit-transform: scale(-1, 1);  -o-transform: scale(-1, 1); -ms-transform: scale(-1, 1);  transform: scale(-1, 1);  }
  video  { -moz-transform: scale(-1, 1);  -webkit-transform: scale(-1, 1);  -o-transform: scale(-1, 1); -ms-transform: scale(-1, 1);  transform: scale(-1, 1);  }
</style>
<div style='position: fixed; top: 10px; width:inherit; text-align: center; z-index: 1; background-color: rgba(255, 255, 255, 0.6);'>
  <font size=4 color=red><p id="infop" align=left></p></font>
</div>

<a-scene embedded artoolkit='sourceType: webcam;' id="thescene">
    <!-- handle hiro marker -->
    <a-marker preset='hiro' id="lysmarker">
        <!-- This is a lysine side chain, starting from the CA atom, which is set to the origin
        All coordinates have been divided by 2,
        so any distance between two atoms must be multiplied by 2 to get the actual distance-->
        <a-sphere id="lys1" position="0.000    0.000    0.000" color="grey" radius="0.4"></a-sphere>
        <a-sphere id="lys2" position="-0.6035  -0.3175  -0.347" color="grey" radius="0.4"></a-sphere>
        <a-sphere id="lys3" position="-1.2480  -0.037   -0.0455" color="grey" radius="0.4"></a-sphere>
        <a-sphere id="lys4" position="-1.851   -0.3545  -0.3925" color="grey" radius="0.4"></a-sphere>
        <a-sphere id="lys5" position="-2.496   -0.0745  -0.0905" color="grey" radius="0.4"></a-sphere>
        <a-sphere id="lys6" position="-3.075   -0.3795  -0.4235" color="blue" radius="0.4"></a-sphere>
        <a-sphere id="lys7" position="-0.5875  -0.2035  -0.88" color="white" radius="0.2"></a-sphere>
        O
        O
        O
    </a-marker>

    <!-- handle kanji marker -->
    <a-marker preset='kanji' id="glumarker">
        <!-- This is a glutamate side chain, starting from the CA atom, which is set to the origin-->
        <a-sphere id="glu1" position="0.000    0.000    0.000" color="grey" radius="0.4"></a-sphere>
        <a-sphere id="glu2" position="-0.6365  -0.3515  -0.238" color="grey" radius="0.4"></a-sphere>
        <a-sphere id="glu3" position="-1.2485  0.0445  -0.006" color="grey" radius="0.4"></a-sphere>
        <a-sphere id="glu4" position="-1.8755  -0.3015  -0.2405" color="grey" radius="0.4"></a-sphere>
        <a-sphere id="glu5" position="-1.836   -0.8175  -0.563" color="red" radius="0.4"></a-sphere>
        O
        O
        O
    </a-scene>
</body>

<script type="text/javascript">
    var lysopaq = new Array(17)
    var gluopaq = new Array(10)
    setInterval(function(){
        document.getElementById("thescene").object3D.updateMatrixWorld();
        var p1 = new THREE.Vector3(); p1.setFromMatrixPosition(document.getElementById("lys6").object3D.matrixWorld);
        var p2 = new THREE.Vector3(); p2.setFromMatrixPosition(document.getElementById("glu5").object3D.matrixWorld);
        var p3 = new THREE.Vector3(); p3.setFromMatrixPosition(document.getElementById("glu6").object3D.matrixWorld);
        distNO1 = 2 * Math.sqrt( Math.pow(p2.x-p1.x,2) + Math.pow(p2.y-p1.y,2) + Math.pow(p2.z-p1.z,2) )
        distNO2 = 2 * Math.sqrt( Math.pow(p3.x-p1.x,2) + Math.pow(p3.y-p1.y,2) + Math.pow(p3.z-p1.z,2) )
        document.getElementById("infop").innerHTML="<p align='left'>N-O1 distance: " + distNO1 + " Å<br>N-O2 distance: " + distNO2 + " Å<br>Electrostatics: "
        + (-1/distNO1/distNO1 - 1/distNO2/distNO2) + "</p>";

        for (i=0;i<17;i++) {lysopaq[i]=0}
        for (i=0;i<10;i++) {gluopaq[i]=0}

        for (i=1;i<18;i++) {
            var p1 = new THREE.Vector3(); p1.setFromMatrixPosition(document.getElementById("lys"+i).object3D.matrixWorld);
            for (j=1;j<11;j++) {
                var p2 = new THREE.Vector3(); p2.setFromMatrixPosition(document.getElementById("glu"+j).object3D.matrixWorld);
                distancia = 2 * Math.sqrt( Math.pow(p2.x-p1.x,2) + Math.pow(p2.y-p1.y,2) + Math.pow(p2.z-p1.z,2) );
                if (distancia < 3) {
                    lysopaq[i-1]=0.5
                    gluopaq[j-1]=0.5
                }
            }
        }

        for (i=18;i<35;i++) { document.getElementById("lys"+i).setAttribute('opacity',lysopaq[i-18])}
        for (i=11;i<21;i++) { document.getElementById("glu"+i).setAttribute('opacity',gluopaq[i-11])}

    }, 200);
</script>

</html>
```

**Figure S1.** Core code for example B from Table 2 (shown in Figure 1). The red line indicates this is one full continuous line. The three vertical purple circles indicate <a-sphere> tags that were hidden for the sake of simplicity. The full code with all <a-sphere> tags is available at https://lucianoabriata.altervista.org/jsinscience/arjs/armodeling/smallmolclashdetection.html

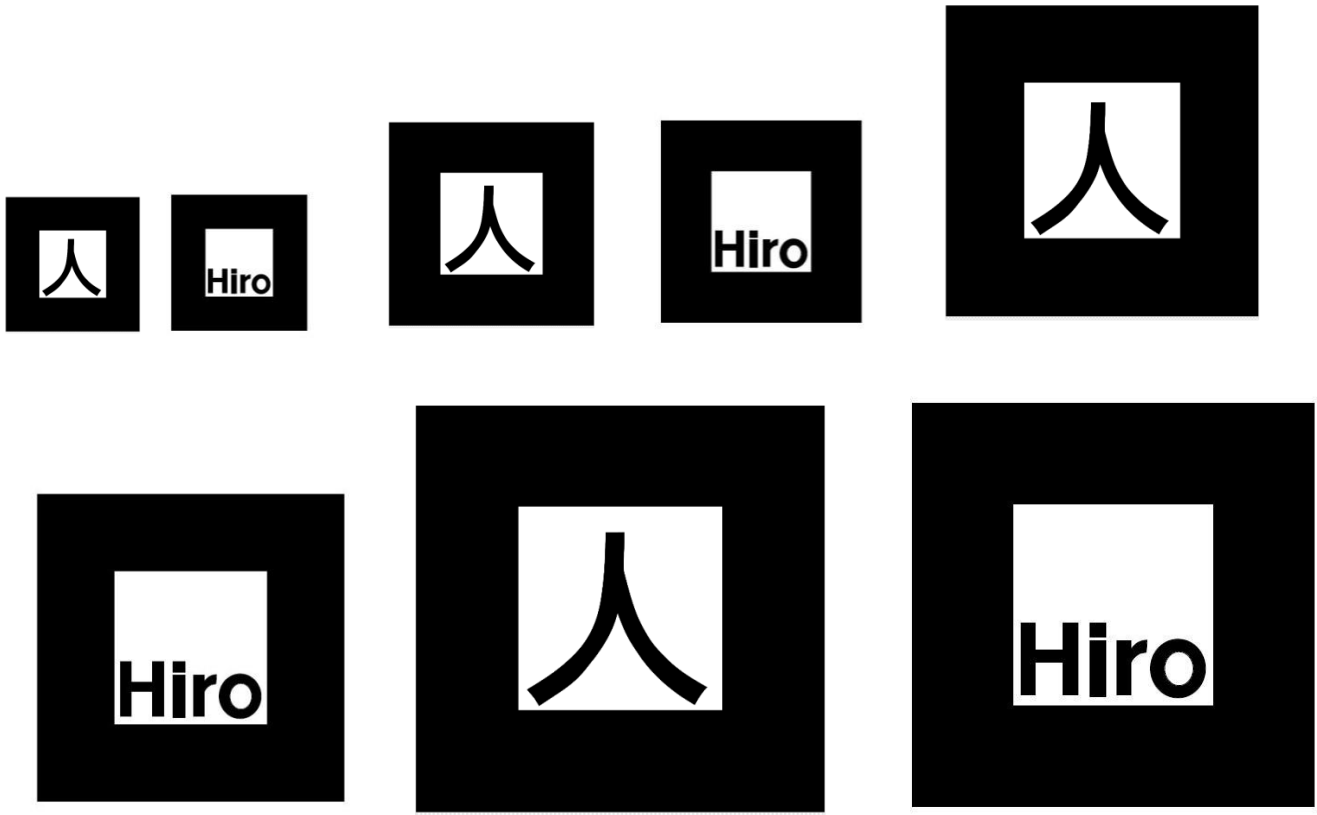

**Figure S2.** The Kanji and Hiro markers laid out in four different sizes, ready to print. It is recommended to glue the used markers on small devices that can be easily handled.

| Laptop computer with integrated webcam | Computer with webcam pointing down |
| *Handheld configuration from front* | *Handheld configuration from top* |

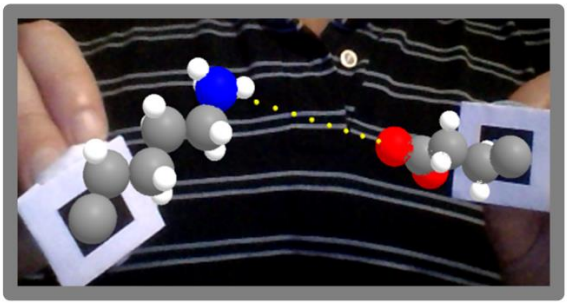

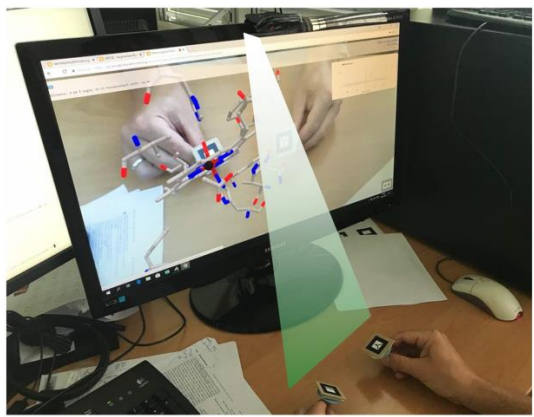

| Large computing power on device, yet easy to carry around | Large computing power on device More relaxed arms |

| Tablet or Smartphone | Smartphone + Cardboard |
| *See-through* | *Head-mounted configuration* |

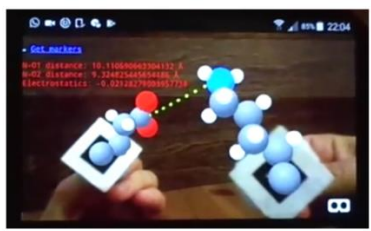

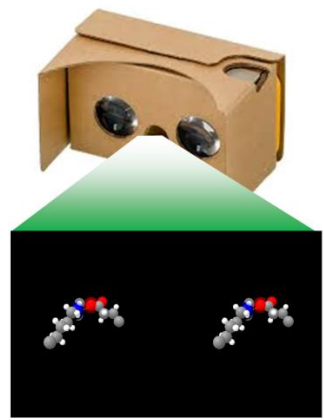

| More relaxed arms and larger screen than in smartphones | More immersive |

**Figure S3.** Different hardware layouts for the proposed web apps, and their key advantages.

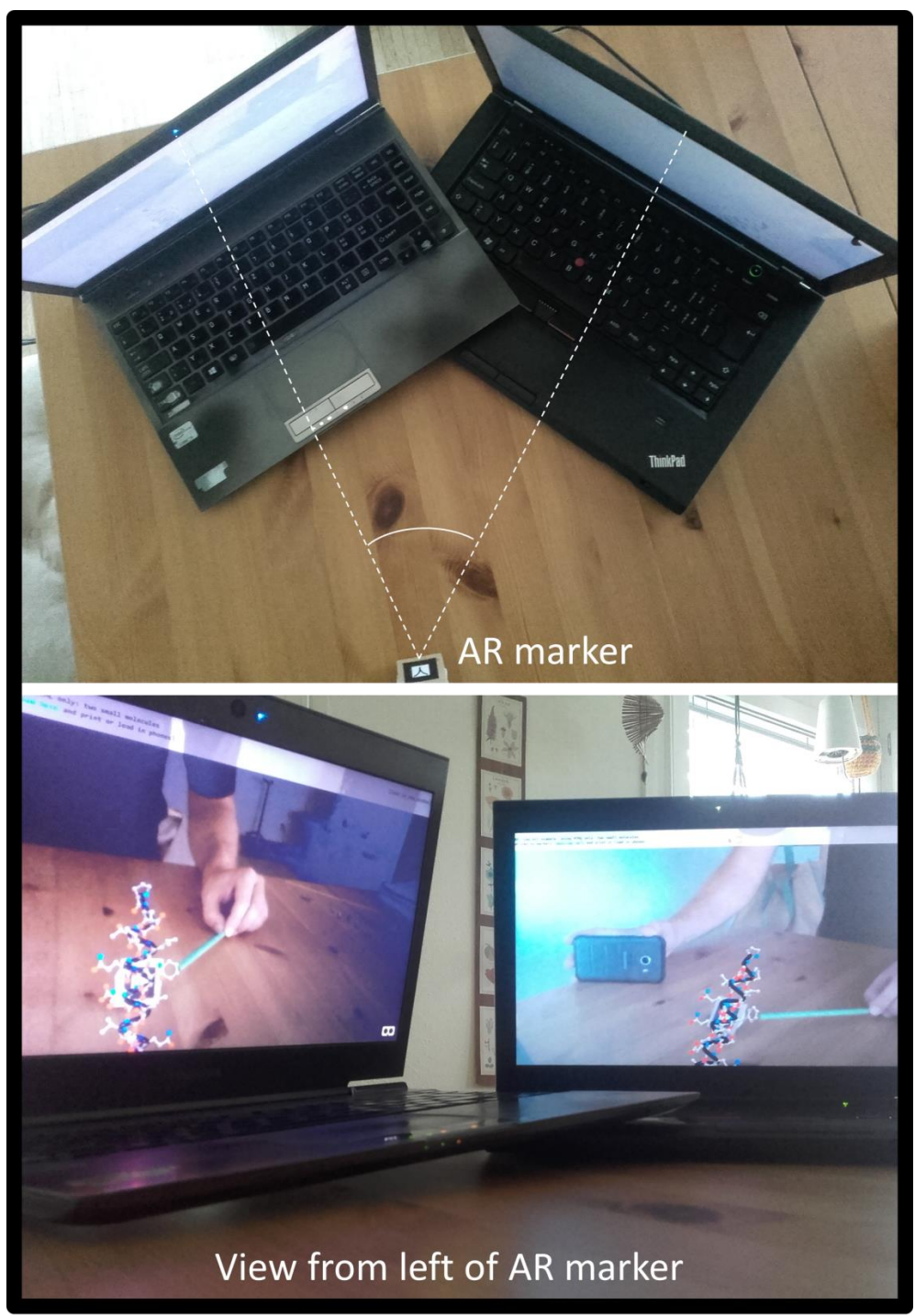

**Figure S4.** The same marker used for augmented reality in two laptops set at around 30º (top). When a user "touches" one residue (a phenylalanine's aromatic ring in the example) other users see the same from their viewpoints. This is the principle for concurrent, multiuser AR/VR.